\documentclass[aps,physrev,twocolumn,nofootinbib,floatfix]{revtex4-2}

\usepackage{amsmath,amssymb,amsfonts}
\usepackage{graphicx}
\usepackage{bm}
\usepackage{placeins}
\usepackage{etoolbox}
\usepackage{hyperref}
\hypersetup{hidelinks,hypertexnames=false}
\makeatletter
\apptocmd{\thebibliography}{\footnotesize}{}{}
\brokenpenalty\@M
\makeatother

\newcommand{\kbt}{k_{\rm B}T}
\newcommand{\eunit}{E_0}
\newcommand{\avgQ}{\langle Q\rangle}
\newcommand{\Vth}{V_{\bm{\theta}}}
\newcommand{\parth}{\bm{\theta}}
\newcommand{\xv}{\mathbf{x}}
\newcommand{\etal}{\textit{et al.}}

\begin{document}

\def\andname{\unskip,}

\title{Online local learning for generative thermodynamic computing}

\author{Huilin Wang}
\email[Contact author: ]{wanghuilin@mails.ccnu.edu.cn}
\affiliation{Key Laboratory of Quark and Lepton Physics (MOE) and
Institute of Particle Physics, Central China Normal University,
Wuhan 430079, China}

\author{Weibing~Deng}
\email[Contact author: ]{wdeng@mail.ccnu.edu.cn}
\affiliation{Key Laboratory of Quark and Lepton Physics (MOE) and
Institute of Particle Physics, Central China Normal University,
Wuhan 430079, China}

\begin{abstract}
Generative thermodynamic computers turn thermal noise into structured
data through Langevin dynamics. We train these systems with a local
update at each integration step. The reverse-path Onsager--Machlup
objective yields a coupling gradient that is a symmetric sum of local
residual--state correlations. We apply this gradient immediately rather
than accumulating it over a full trajectory. In digital simulations
using MNIST prototypes, online and trajectory-batch training reach
similar validation losses on fixed noising paths. Models trained online
release less heat on average in all five independently seeded
pairs, with both models' parameters held fixed during sampling.
Auxiliary classifier and nearest-prototype measures change modestly,
while pairwise diversity decreases. The response to noise depends
strongly on where the errors enter: independent zero-mean errors in the
formed updates produce little heat change over a finite range of noise
amplitudes, whereas residual offset and temporal correlation have much
larger effects. Storing trained couplings requires substantially less
precision than resolving deterministic updates during training.
Together, these results establish a local online training method and
show how update timing, noise structure, and precision affect
generative thermodynamic computing.
\end{abstract}

\maketitle
\raggedbottom

\section{Introduction}
\label{sec:intro}

In thermodynamic computing, fluctuating physical variables perform a
calculation through their collective
dynamics~\cite{Conte2019,Hylton2020,Wimsatt2021,Boyd2022}.
A Langevin system, for example, can encode a calculation in its
relaxation or in a nonequilibrium trajectory.
This approach draws on stochastic
thermodynamics~\cite{Seifert2012,Jarzynski1997,Crooks1999} and is motivated by the growing energy
cost of machine-learning workloads: conventional logic remains far
above the Landauer scale, and data movement and arithmetic dominate
modern accelerators~\cite{Landauer1961,Berut2012,Horowitz2014}.

Thermodynamic computers have been used for a range of tasks.
Equilibrium devices can solve linear-algebra problems such
as matrix inversion~\cite{Aifer2024,Melanson2025}; nonlinear
out-of-equilibrium devices can perform machine-learning calculations
analogous to those of neural networks~\cite{WhitelamCasert2024}; and
stochastic $p$-bit devices can solve optimization problems
natively~\cite{Camsari2017,Borders2019}.
Gradient descent extends this approach to tasks learned from data.
Whitelam~\cite{Whitelam2025gen} showed that a generative thermodynamic
computer trained by maximizing the probability of reversing a noising
trajectory learns to synthesize structured outputs from thermal noise.
The reverse-path formulation connects generative learning with
stochastic thermodynamics and provides a physical counterpart to
score-based generative models in machine
learning~\cite{SongErmon2019,Ho2020,SohlDickstein2015}.
Whitelam~\cite{Whitelam2025gd} subsequently demonstrated that gradient
descent can also train a thermodynamic computer to mimic a neural
network trained for image classification and analyzed its modeled
energetic cost relative to digital implementations.
A distinct hybrid approach, thermodynamic natural gradient descent,
uses an equilibrium analog subsystem to accelerate second-order
optimization of conventionally represented models~\cite{Donatella2026TNGD}.
Here the Langevin system is itself the nonequilibrium generator, and
its couplings are updated from signals measured at each integration step.

Can the computer learn one step at a time? The analytic gradient is a
sum of local contributions from individual integration steps. Applying
each contribution immediately gives an online rule whose behavior can
differ from trajectory-batch training at finite learning rate.
In the generative Langevin-computer framework of
Ref.~\cite{Whitelam2025gen}, the demonstrated training procedure
integrates the Langevin dynamics, accumulates gradients over a complete
noising trajectory, and applies a batch parameter update at the end.
This requires a gradient accumulator for each parameter, but no
storage of the trajectory states.
An \emph{online} rule instead updates the couplings at each integration
step from local information. Each update changes the parameters used
to evaluate the next gradient.
The gradient of the Onsager--Machlup loss with respect to a coupling
$J_{ij}$ requires only the endpoint states, displacements, and local
forces of nodes $i$ and $j$ at a single timestep.
We write the analytic gradient of Ref.~\cite{Whitelam2025gen} as a
symmetric sum of local residual--state correlations.
We then derive the finite-step difference between online and batch
training and test its consequences numerically.
The use of local measurements connects this rule to physical learning
in decentralized networks~\cite{Dillavou2022}, equilibrium
propagation~\cite{Scellier2017}, and coupled learning
networks~\cite{Stern2021,Lopez-Pastor2023}.
A closely related residual-driven framework is the Hebbian Physics
Network of Auti \etal~\cite{Auti2026HPN}, in which local violations of
transport laws drive adaptation of a constitutive operator for diffusion
and flow problems. In our setting, the residual comes from the
Onsager--Machlup loss for one reverse step of a stochastic generative
process. We study how stepwise and trajectory-batch updates differ at
finite learning rate, and how the resulting models differ in the heat
released during generation with fixed parameters.

Local learning has also been investigated recently in other physical
generative architectures.
B{\"o}sch \etal~\cite{Bosch2025Local} derived local rules for learning
time-dependent driving protocols in out-of-equilibrium score-based
physical generative models, using either force measurements or observed
dynamics, and demonstrated generation in nonlinear oscillator networks.
Yu \etal~\cite{Yu2025NeuralLangevin} introduced a neural Langevin
machine with a local asymmetric plasticity rule based on recurrent
network fixed points and Langevin sampling.
We use the autonomous generative thermodynamic computer of
Ref.~\cite{Whitelam2025gen} and train its symmetric pair couplings using
the reverse-path Onsager--Machlup objective. The learned couplings
remain fixed when the model generates new samples.

A second question concerns the robustness of this local rule.
Errors can enter after an edge update has been formed, or earlier in
the node residual used to construct it. Limited precision can likewise
affect either the storage of trained couplings or the small increments
applied during training. These distinctions matter both for the
stability of the numerical method and for future physical implementations.

We compare online and batch training across independent seeds and find
similar validation losses but lower generation heat for models trained
online. We examine this difference by varying the learning rate and
update timing, then decomposing the energy at the generated endpoints.
Noise and quantization tests show how the response depends on error
structure and distinguish the precision needed to store couplings from
that needed to update them.

\section{Model and Training Rules}
\label{sec:model}

\subsection{Thermodynamic computer}

We follow the model of Refs.~\cite{Whitelam2025gen,Whitelam2025gd}.
The computer comprises $N = N_v + N_h$ classical real-valued degrees
of freedom $\xv = \{x_i\}$, split into $N_v = 784$ visible units
(the $28\times 28$ display layer) and $N_h = 512$ hidden units.
Couplings $J_{ij}$ exist between every visible--hidden pair and
every hidden--hidden pair; visible units have no direct mutual
couplings, following Ref.~\cite{Whitelam2025gen}.
The hidden--hidden diagonal is fixed to zero after every update.
The units obey the overdamped Langevin equation
\begin{equation}
  \dot{x}_i = -\mu\,\partial_i \Vth(\xv)
             + \sqrt{2\mu\kbt}\;\eta_i(t),
  \label{eq:langevin}
\end{equation}
where $\mu$ is the mobility, $\kbt$ is the thermal energy, and
$\eta_i(t)$ is unit-variance Gaussian white noise.
The potential energy is
\begin{equation}
  \Vth(\xv) = \sum_{i=1}^N \!\bigl(J_2 x_i^2 + J_4 x_i^4\bigr)
            + \sum_{a\in h} b_a x_a
            + \sum_{(ij)} J_{ij} x_i x_j,
  \label{eq:potential}
\end{equation}
where the sum $\sum_{(ij)}$ runs over all connected pairs with
the convention $J_{ij} = J_{ji}$ (symmetric couplings),
fixed on-site parameters $J_2 = J_4 = \eunit$, and trainable
parameters $\parth = (J_{vh},J_{hh},b_h)$.
Only hidden-unit biases are trained; visible-unit model biases are fixed
to zero.  All trainable couplings and hidden biases are initialized to
zero in every run.
This energy-based architecture is closely related to classical
Hopfield networks~\cite{Hopfield1982} and Boltzmann
machines~\cite{Ackley1985}, but operates in continuous
state space with nonequilibrium Langevin dynamics.
The quartic term ($J_4>0$) provides nonlinearity and keeps the onsite
potential confining at large amplitude~\cite{WhitelamCasert2024}.
We integrate Eq.~\eqref{eq:langevin} with the Euler--Maruyama scheme
using single-precision state and parameter arrays.  We choose the
simulation energy unit $\eunit$ such that
$J_2 = J_4 = \eunit = 1$ and set $\kbt = 0.1\,\eunit$,
$\mu = 1$, trajectory time $t_f = 2.5\,\mu^{-1}$,
integration timestep $\Delta t = 10^{-3}$,
and learning rate $\alpha = \Delta t / t_f = 4\times 10^{-4}$.

\subsection{Batch training (reference)}

Training follows Ref.~\cite{Whitelam2025gen}.
Each prototype $s\in\mathbb{R}^{784}$ is normalized to zero mean and
unit variance.
A fixed random projection $P\in\mathbb{R}^{784\times512}$, drawn once
with entries $P_{ia}\sim\mathcal{N}(0,1/784)$ using seed 42, supplies
the corresponding hidden-layer field.
Starting from zero, the uncoupled units are first evolved for $t_f$
under external forces $2s$ and $2P^{\mathsf T}s$ on the visible and
hidden units, respectively.
A noising trajectory $\omega=\{\xv(t_k)\}_{k=0}^K$ is then generated
with external force
$(a(t)s,a(t)P^{\mathsf T}s)$, where
$a(t)=2[1-t/(0.75t_f)]$ for $t<0.75t_f$ and $a(t)=0$ thereafter.
The learned couplings and hidden biases do not drive this forward
noising path; they enter the candidate reverse-path loss below.
Training maximizes the probability of the reverse trajectory
$\tilde\omega$ under the candidate model.
From the discrete Onsager--Machlup
action~\cite{OnsagerMachlup,Cugliandolo2017}, the loss for one
reverse step is
\begin{equation}
  \mathcal{L}_{\rm step}^{(k)}
  = \sum_{i=1}^N
    \frac{\bigl(-\Delta x_i + \mu\,\partial_i\Vth(\xv')\Delta t\bigr)^2}
         {4\mu\kbt\,\Delta t},
  \label{eq:loss_step}
\end{equation}
where $\xv' = \xv(t_{k+1})$ and $\Delta\xv = \xv'-\xv(t_k)$.
The total loss is $\mathcal{L} = \sum_k \mathcal{L}_{\rm step}^{(k)}$.
For plots and validation on fixed paths, we report the loss per step,
$\overline{\mathcal{L}}=\mathcal{L}/K$. The update equations below use
the sum over the full trajectory.
The \emph{batch} update accumulates gradients over the full trajectory
before updating parameters once:
\begin{equation}
  \parth \;\leftarrow\; \parth
    - \alpha \sum_{k=1}^K \nabla_{\parth}\,
    \mathcal{L}_{\rm step}^{(k)}.
  \label{eq:batch_update}
\end{equation}
We use the analytic gradients of Ref.~\cite{Whitelam2025gen} to
construct the online rule below.

\subsection{Online local learning rule}

Each symmetric coupling ($J_{ij} = J_{ji}$) contributes
$J_{ij} x_i x_j$ to the potential and therefore enters the forces on
both connected units. Differentiating Eq.~\eqref{eq:loss_step} with
respect to $J_{ij}$ gives
\begin{equation}
  \frac{\partial \mathcal{L}_{\rm step}}{\partial J_{ij}}
  = r_i\, x_j' + r_j\, x_i',
  \label{eq:grad_Jij}
\end{equation}
where
\begin{equation}
  r_i = \frac{-\Delta x_i + \mu\,\partial_i\Vth(\xv')\Delta t}
             {2\kbt}
  \label{eq:residual}
\end{equation}
is the \emph{local prediction residual} of unit $i$. It measures the
scaled difference between the reverse of the observed displacement
$\Delta x_i$ and the displacement predicted by the current potential
for that reverse step.
It depends on the state and displacement of unit $i$ and on the local
force $\partial_i\Vth$, which includes its neighbors' states.
The coupling gradient is thus a symmetric sum of two local
residual--state correlations, $r_i x_j'$ and $r_j x_i'$. Both can be
formed at the edge connecting the units. The batch rule accumulates
the same terms without storing the path.

The \emph{online} rule applies these gradients immediately at each
integration step $t_k$,
\begin{align}
  J_{ij}(t_{k+1}) &= J_{ij}(t_k)
    - \alpha \bigl(r_i x_j' + r_j x_i'\bigr),
  \label{eq:online_J} \\
  b_a(t_{k+1}) &= b_a(t_k) - \alpha\, r_a,
  \label{eq:online_b}
\end{align}
where the bias update is applied only to hidden units $a\in h$.
Each update uses the current displacement, local force, and neighboring
states. The trajectory-level gradient accumulator is no longer needed.

To see how online and batch training differ at finite learning rate, let
$g_k(\parth)=\nabla_{\parth}\mathcal{L}_{\rm step}^{(k)}(\parth)$
denote the step-$k$ gradient along a fixed noising trajectory.
The batch update from initial parameters $\parth_0$ is
\begin{equation}
  \parth_{\rm batch}
  = \parth_0 - \alpha\sum_k g_k(\parth_0).
\end{equation}
The online rule instead evaluates later gradients at already updated
parameters:
\begin{align}
  \parth_{\rm online}
  &= \parth_0 - \alpha\sum_k g_k(\parth_{k-1}) \nonumber\\
  &= \parth_{\rm batch}
     + \alpha^2\sum_{k>l} H_k(\parth_0)g_l(\parth_0)
     + \mathcal{O}(\alpha^3),
  \label{eq:online_batch_expansion}
\end{align}
where $H_k$ is the Jacobian of $g_k$, assuming smooth gradients and
sufficiently small $\alpha$.
Thus batch and online training agree to first order in $\alpha$ but
differ at second and higher orders.
They can therefore reach different parameter sets even when their
loss curves are similar.

\subsection{Noise and coupling quantization}

We test noise in the local update signals and finite precision in the
couplings.

\textbf{Update and residual noise.}
We introduce errors at two stages of the local update calculation.
In the formed-update model, independent Gaussian errors are added
after constructing each coupling or hidden-bias update signal,
$u_{ij}\to u_{ij}+\sigma_{\rm upd}\zeta_{ij}$ and
$u_a\to u_a+\sigma_{\rm upd}\zeta_a$, where
$u_{ij}=r_i x_j'+r_j x_i'$ and $u_a=r_a$ is the hidden-bias update.
For $J_{hh}$, one $\zeta_{ij}$ is drawn per undirected edge and mirrored
so that coupling symmetry is preserved.
In the residual model, noise is added before forming the edge update,
\begin{equation}
  r_i \;\to\; r_i + \sigma_{\rm res}\,\xi_i,\qquad
  \xi_i \sim \mathcal{N}(0,1),
  \label{eq:noise}
\end{equation}
with $\sigma_{\rm upd},\sigma_{\rm res}\geq0$.
The two amplitudes describe errors in different variables. A
quantitative comparison therefore requires a mapping between the noise
channels.
We additionally test a residual offset $\mu_{\rm bias}$ and an AR(1)
residual error
$\xi_{k+1}=\rho\xi_k+\sqrt{1-\rho^2}\,\epsilon_k$, reset at the start of
each training trajectory with $\xi_0\sim\mathcal{N}(0,1)$.
The offset screen adds the same $\mu_{\rm bias}$ to every node residual
on top of iid residual noise; all injected-error random streams are
independent of the thermal Langevin random stream.
Within each response sweep, the thermal training seed is fixed across
error amplitudes, and all conditions use a common generation seed.

\textbf{Coupling quantization.}
After training with full (32-bit) precision, the stored coupling tensors
$A\in\{J_{vh},J_{hh}\}$ are quantized to $b_{\rm storage}$ bits; the
primary screen retains $b_h$ at full precision.  We write $b$ for
$b_{\rm storage}$ in Eqs.~\eqref{eq:quant_n} and~\eqref{eq:quant}.
Here ``32 bit'' denotes the unquantized single-precision simulation
reference rather than a 32-bit fixed-point grid.
For each tensor separately, an entry is mapped to the
nearest integer level:
\begin{equation}
  \hat{n}_{ij} = \text{round}\!\bigl(
    A_{ij}/A_{\max}\cdot(2^{b-1}-1)\bigr),
  \label{eq:quant_n}
\end{equation}
where $A_{\max}=\max_{ij}|A_{ij}|$ and
$\hat{n}_{ij} \in \{-(2^{b-1}-1),\ldots,+(2^{b-1}-1)\}$.
The quantized coupling is then
\begin{equation}
  A_{ij}^{(b)} = \frac{\hat{n}_{ij}}{2^{b-1}-1}\,A_{\max}.
  \label{eq:quant}
\end{equation}
Hidden--hidden couplings are quantized once per undirected edge and
mirrored.
We generate samples with these quantized couplings, leaving the trained
hidden biases unchanged.
For each trained model, we use the same generation random numbers at
all storage bit depths. We compare the resulting heat distributions
with the 32-bit reference using the empirical two-sample
Kolmogorov--Smirnov distance. Because the random numbers pair the
samples across bit depths, the usual independent-sample KS $p$ value
does not apply.
We separately test a training-time precision $b_{\rm train}$: each
online update is followed immediately by clipping and quantization of
the updated $J_{vh}$ and $J_{hh}$ tensors, while $b_h$ remains at full
precision in the primary screen.  The fixed tensor ranges are measured
from the corresponding full-precision checkpoint.
For stochastic rounding, a scaled value $y$ is mapped to
$\lfloor y\rfloor+1$ with probability $y-\lfloor y\rfloor$ and to
$\lfloor y\rfloor$ otherwise, using a random stream independent of the
thermal dynamics.
Training-time precision conditions use the same thermal training and
generation seeds as the full-precision reference.

\subsection{Training versus generation}

Online updates are applied only along the noising trajectories used
for training.
For each generation sample, all states are reset to zero and evolved
for $t_f$ under the uncoupled onsite potential before the learned model
is applied for a further time $t_f$.
The couplings and biases remain fixed throughout generation and all
subsequent quality and heat measurements.
We measure the heat released as this trained system relaxes,
$Q = V_{\parth}(\xv_0)-V_{\parth}(\xv_{t_f})$, with heat flowing to the
environment taken as positive. The measurement begins after preparation
of the uncoupled initial state and establishment of the trained
parameters; the work of establishing or switching those parameters
lies outside this interval.

In each paired comparison, batch and online training use the same seed
for thermal noise and therefore the same sequence of forward noising
trajectories. Across pairs, the training seeds change the trajectory
realizations, while all parameter initializations remain identical.
The five main training seeds are 101--105.
Batch and online models within a trained pair are also evaluated with
the same generation seed (common random numbers), while distinct
trained pairs use generation seeds 90101--90105.
The fixed validation trajectory uses seed 777001, starts from the first
prototype in training order (digit 0), and is evaluated every 10
training cycles.

To compare the learned potentials, we evaluate all ten final checkpoints
on 30 additional noising paths, ten per training prototype, shared
across checkpoints. We also compare the normalized losses after
subtracting the parameter-independent squared-increment term in
Eq.~\eqref{eq:loss_step}. For the energy diagnostic, we replay the
original 200 generation trajectories per checkpoint and retain the
full visible and hidden endpoint states. The replayed images and heats
match the archived values exactly, and all trainable parameter arrays
are verified to remain unchanged. We evaluate both learned potentials
on both endpoint ensembles. Evaluating states in the other model's
potential diagnoses changes in the energy drop; only evaluation in
their own potential measures the heat released during generation.

\subsection{Auxiliary generation-quality metrics}

We assess generation quality with a separately trained MNIST
classifier, which plays no role in training the thermodynamic computer.
The classifier contains two $3\times3$ convolutional layers (16 and 32
channels), each followed by ReLU and $2\times2$ max pooling, then a
128-unit fully connected layer and a 10-class output.
It is trained with Adam (learning rate $10^{-3}$) for five epochs using
seed 42.
For classifier input only, each generated image is linearly rescaled by
its 2nd and 98th percentiles to $[0,1]$, followed by standard MNIST
normalization.
We report the mean maximum softmax score, the fraction assigned to the
three training classes, and the natural-log entropy of the
predicted-class counts. These labels and scores give auxiliary measures
of recognizability. Their interpretation depends on preprocessing,
because the signed, zero-mean generated states differ from the standard
MNIST images used to train the classifier. The maximum softmax score
is not a calibrated probability or a confidence interval.
Nearest-prototype distance is the Euclidean distance between an
unrescaled generated state and the closest zero-mean/unit-variance
training prototype.
Pairwise diversity is computed directly from unrescaled generated
states and averaged over all distinct sample pairs.
For statistical comparisons of batch and online training, each trained
seed pair is one independent observation.
We report paired seed-level differences with two-sided Student-$t$
confidence intervals and tests, together with the exact two-sided sign
test for the heat comparison. The spread across trajectories from one
trained model describes sampling variability, not independent
replication of the training algorithm.

\section{Results and Discussion}
\label{sec:results}
\label{sec:discussion}

The main comparison uses three MNIST digits
(0, 1, 2)~\cite{LeCun1998}. We take the first image of each class in the
training set, cycle through them in the order (0, 1, 2), and normalize
each image to zero mean and unit variance.
Using one image per class follows Ref.~\cite{Whitelam2025gen}, where a
small set of prototypes was sufficient to train a Langevin computer to
generate structured outputs.
For this comparison, training runs for $N_{\rm cycle} = 300$ cycles.
We measure the mean heat emitted per denoising trajectory after the
trained parameters are frozen,
$\avgQ = \langle V_{\parth}(\xv_0) - V_{\parth}(\xv_{t_f})\rangle$,
averaged over $N_{\rm samp} = 200$ independent runs, each prepared by
evolving the uncoupled units for the same finite time.
Sample quality, class coverage, and diversity provide complementary
measures of generation. The 2-bit quantization result below illustrates
why heat alone does not determine fidelity.
We express heat in the simulation energy unit $\eunit$.
Since $\kbt = 0.1\,\eunit$, conversion to thermal units multiplies the
reported values by ten. This conversion leaves relative heat differences
and the dimensionless perturbation coordinates unchanged.

\begin{figure*}[t]
  \centering
  \includegraphics[width=\textwidth]{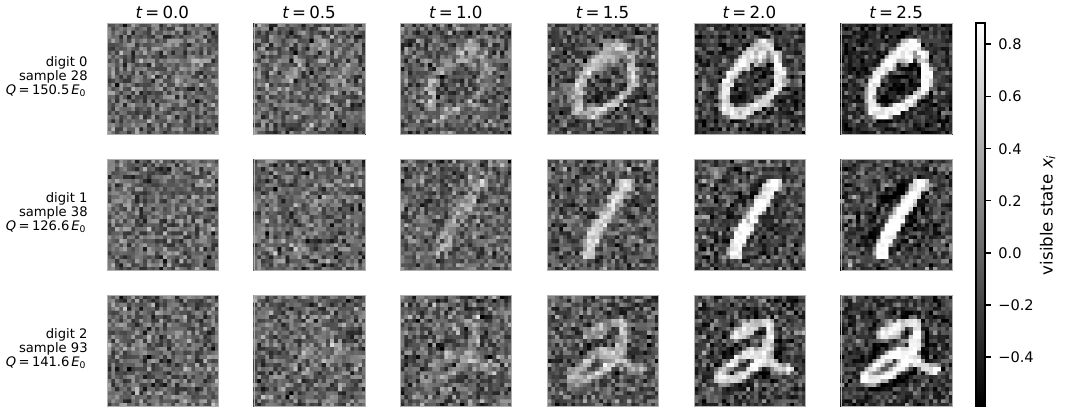}
  \caption{%
    \textbf{Fixed-seed denoising trajectories for digits 0, 1, and 2.}
    For each target digit, the displayed trajectory has the minimum
    nearest-prototype distance among the 200 archived online seed-101
    samples assigned to that digit by the frozen classifier (archive
    samples 28, 38, and 93).  Full-archive diagnostics appear in
    Fig.~\ref{fig:samples}(b--d).
    Each row shows six equally spaced times from generation seed 90101.
    Terminal states and heats were verified against the saved archive,
    and all states share one grayscale set by their 1st and 99th
    percentiles.%
  }
  \label{fig:overview}
\end{figure*}

Figure~\ref{fig:overview} illustrates the denoising process for the
three target digits.
Starting from thermal noise at $t=0$, the coupled Langevin dynamics
progressively transforms the noise into structured digit-like images
by $t = t_f$.
Thermal fluctuations produce different trajectories and variation among
the generated images.
To display the learned structures, we select the archived sample with
the smallest nearest-prototype distance within each classifier-assigned
target class. Figure~\ref{fig:samples}(b--d) gives the class counts and
quality measures for the complete archive.

\subsection{Generation heat and sample quality}
\label{sec:batch_vs_online}

\begin{figure*}[t]
  \centering
  \includegraphics[width=\textwidth]{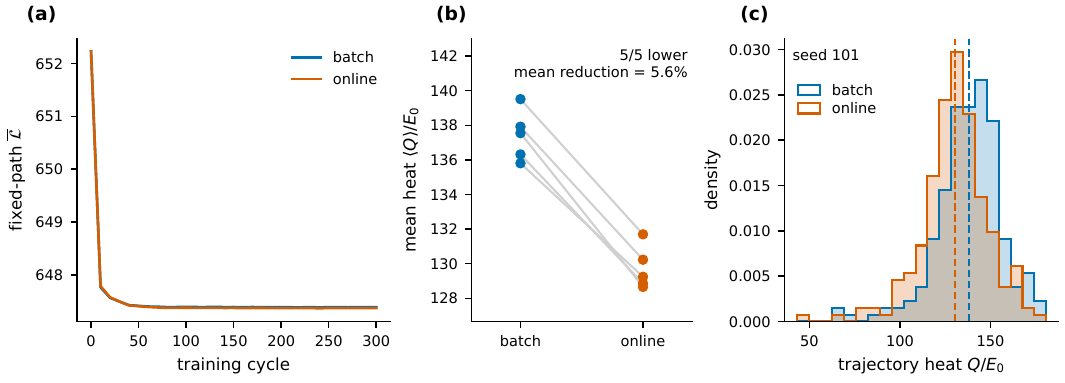}
  \caption{%
    \textbf{Paired comparison of online and batch training.}
    (a)~Mean fixed-path validation loss across five training seeds;
    shading denotes 1 s.d. across seeds.
    (b)~Seed-level mean heat, with each line joining one paired batch and
    online run. Online training is lower in all five pairs, with a mean
    relative reduction of $5.6\%$.
    The paired $t$-test gives $p=3.3\times10^{-5}$; the exact two-sided
    sign test gives $p=0.0625$.
    (c)~Trajectory-level heat distributions for trained pair 101
    ($N_{\rm samp}=200$ per model); dashed lines mark the means.
    Panel (c) describes one trained pair; panel (b) compares independent
    trained seed pairs.%
  }
  \label{fig:batch_vs_online}
\end{figure*}

Figure~\ref{fig:batch_vs_online} compares the batch and online
learning rules.
Both rules approach the same validation-loss plateau within approximately
50 training cycles and remain stable for the full training run
[Fig.~\ref{fig:batch_vs_online}(a)].
The validation curves track one another closely on the fixed forward
path. We measure generation heat separately, on denoising trajectories
of the frozen trained models.

Despite their similar validation losses, the trained models release
different amounts of heat. We compare five independently seeded pairs,
treating each pair as one statistical unit. The trajectory distributions
for seed 101 illustrate the difference within one pair
[Fig.~\ref{fig:batch_vs_online}(c)].
The online computer emits less heat in all five paired seeds, with a
mean relative reduction of $5.6\%$ [Fig.~\ref{fig:batch_vs_online}(b)].
The mean paired absolute reduction is
$7.68\,\eunit$ (95\% Student-$t$ confidence interval
$[6.64,8.71]\,\eunit$; paired $t$-test
$p=3.3\times10^{-5}$; exact two-sided sign test $p=0.0625$ for
$n=5$).
All pairs show the same trend, although the small number of pairs
limits the power of the nonparametric sign test.
Equation~\eqref{eq:online_batch_expansion} shows that the two rules
follow different finite-step optimization trajectories even when they
agree to first order in the learning rate.
Varying the learning rate and update delay
(Sec.~\ref{sec:numerical_controls}) tests the role of update timing.
We then examine the learned potentials and generated endpoint states
to locate the heat difference (Sec.~\ref{sec:path_heat}).

\begin{figure*}[!t]
  \centering
  \includegraphics[width=\textwidth]{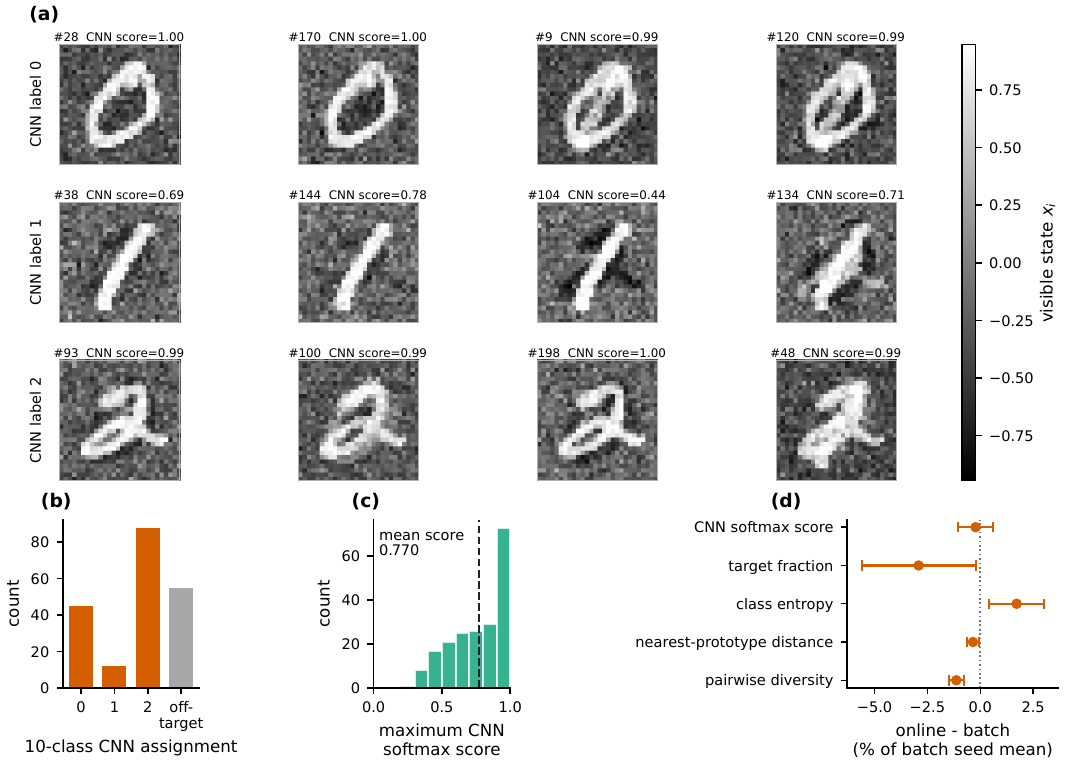}
  \caption{%
    \textbf{Generated samples and auxiliary quality diagnostics.}
    (a)~Class-balanced low-distance display for online-trained seed 101:
    within each classifier-assigned target class, the four archived
    samples with smallest nearest-prototype distance are shown.  Titles
    give archive index and maximum CNN softmax score.  The score is not a
    calibrated probability or confidence interval. This selection shows
    low-distance examples; class frequencies and typical distances are
    assessed from the full archive. A single grayscale, set by the 1st and 99th
    percentiles of all 200 unrescaled samples, is shared by every image;
    samples are not individually contrast normalized.
    (b,c)~Ten-class auxiliary-CNN assignments and maximum softmax scores
    for all 200 samples from this model.  Orange bars identify the three
    classes used to train the thermodynamic computer; the gray bar
    collects CNN assignments outside the training set. The dashed line in
    panel (c) marks the mean score.
    (d)~Five-seed paired online-minus-batch differences with 95\%
    Student-$t$ confidence intervals, expressed as a percentage of each
    metric's batch seed-level mean. Each model contributes 200 stored
    samples. For every model, a separate generation call verified
    elementwise that all parameter arrays remained unchanged.%
  }
  \label{fig:samples}
\end{figure*}

Figure~\ref{fig:samples} shows generated samples drawn from one
online-trained computer and their evaluation by an independent
MNIST classifier (a convolutional neural network trained to $98.7\%$
test accuracy on the standard MNIST dataset).
Panel (a) displays the four samples closest to a training prototype
within each classifier-assigned target class. Panels (b--d) summarize
the complete sample archives.
For seed 101, the CNN assigns most samples to the three training
classes [Fig.~\ref{fig:samples}(b)], predominantly to class 2 and rarely
to class 1. The remaining assignments fall outside the target set.
Most of these off-target samples are closest to a class-1 or class-2
prototype in the unrescaled state space. Nearest-prototype assignment
can only choose among the stored classes, however, so it does not
establish whether an off-target sample is a recognizable target digit.
All off-target assignments are included in the quality analysis.
A 3-output classifier would assign every sample to classes 0, 1, or 2,
concealing this distinction. The ten-class evaluation and the displayed
structures together show digit generation with strongly uneven class
coverage for this seed.

We compare generation quality across the same five trained seed pairs.
For each seed, batch and online models were evaluated with the same
generation random seed, 200 saved samples per model, and frozen parameters.
An additional generation call for each model verified, by elementwise
comparison before and after sampling, that all coupling and bias arrays
remained unchanged.
The paired intervals in Fig.~\ref{fig:samples}(d) resolve no change in
the mean maximum softmax score. Online samples have a lower target-class
fraction, higher class entropy, slightly smaller nearest-prototype
distance, and lower pairwise diversity. Lower heat therefore comes
with a mixed change in quality: the outputs are slightly closer to the
prototypes but less diverse, and the CNN assigns fewer of them to the
target classes.

To test whether the heat trend persists with more prototypes,
we repeated the comparison with the first five training images from each
of the same three classes (15 prototypes in total), using three
independently seeded pairs, 300 training cycles, and 200 generated samples
per model.
Online heat is lower in all three pairs, with a mean relative reduction
of about $7\%$.  Maximum CNN softmax score and target-class fraction
remain similar in this auxiliary screen. The online samples again have
lower pairwise diversity, but their nearest-prototype distance is
slightly larger. Thus the heat and diversity trends persist across
these three seeds, while the change in prototype distance reverses.

\subsection{Perturbation response}
\label{sec:noise}

\begin{figure*}[t]
  \centering
  \includegraphics[width=0.96\textwidth]{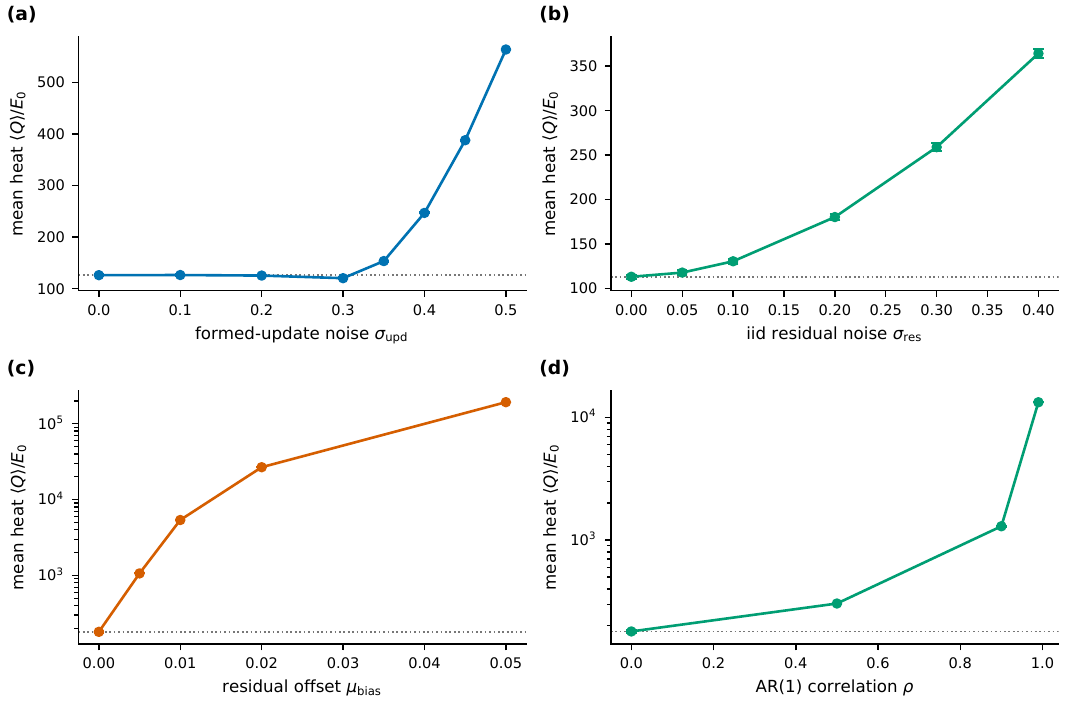}
  \caption{%
    \textbf{Perturbation-response screens.}
    (a)~Zero-mean noise $\sigma_{\rm upd}$ added after each local update
    has been formed ($N_{\rm cycle}=100$, $N_{\rm samp}=200$).
    (b)~Iid noise $\sigma_{\rm res}$ added to node residuals before the
    symmetric edge update is formed.
    (c)~Uniform residual offset $\mu_{\rm bias}$ added on top of iid
    residual noise with $\sigma_{\rm res}=0.2$.
    (d)~AR(1) temporal correlation at the same residual-noise amplitude.
    Panels (b--d) use $N_{\rm cycle}=40$ and $N_{\rm samp}=60$.
    Error bars are trajectory SEM; panels (c,d) use logarithmic vertical
    axes.  Gray dotted lines mark the corresponding zero-perturbation
    baseline in each panel.  The two noise amplitudes in (a,b) act on
    different variables and are not quantitatively interchangeable.
    Each panel reports a single trained-seed response curve.%
  }
  \label{fig:noise_response}
\end{figure*}

Figure~\ref{fig:noise_response}(a,b) distinguishes the two noise channels
defined in Sec.~\ref{sec:model}.
For formed-update noise, heat remains near its baseline through
$\sigma_{\rm upd}=0.30$ before crossing over to large heat degradation
between 0.30 and 0.35. Heat decreases modestly at low noise; sample
quality was not measured in this sweep, so the decrease alone does not
indicate better generation.

Residual noise enters a node signal shared by all of its incident edge
updates. It produces a gradual, monotonic heat increase up to
$\sigma_{\rm res}=0.40$. The noise-free baselines differ because the two
screens use different training lengths and sampling protocols.
The amplitudes $\sigma_{\rm upd}$ and $\sigma_{\rm res}$ label different error
channels, each with its own response curve.

We next add an offset or temporal correlation to the residual noise,
keeping $\sigma_{\rm res}=0.2$
[Fig.~\ref{fig:noise_response}(c,d)].
Even the smallest nonzero offset produces a marked heat increase, and
larger offsets change the heat by orders of magnitude. Increasing
AR(1) correlation also raises the heat, particularly at the two largest
tested correlations. In these tests, coherent offsets and persistent
errors have much larger effects than independent zero-mean errors in
the formed updates.

\subsection{Precision response}
\label{sec:quant}

\begin{figure*}[t]
  \centering
  \includegraphics[width=0.96\textwidth]{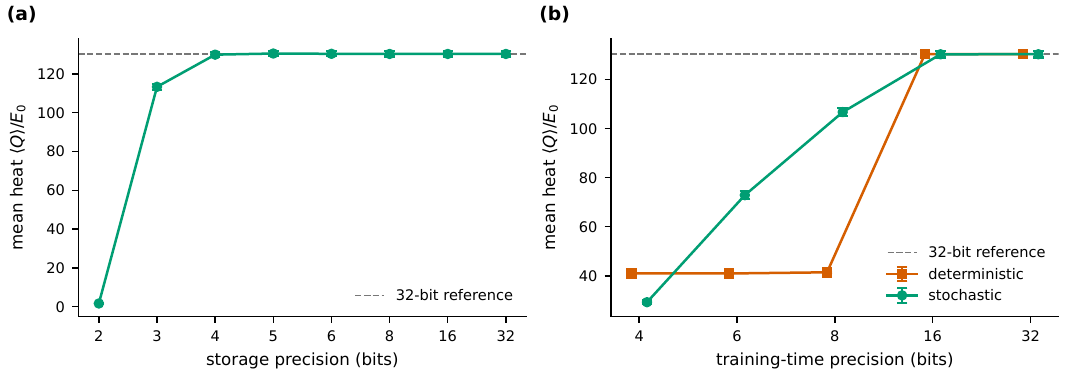}
  \caption{%
    \textbf{Coupling-precision screens.}
    (a)~Post-training quantization of $J_{vh}$ and $J_{hh}$, with $b_h$
    retained at full precision.
    (b)~Deterministic and stochastic quantization of the couplings after
    every online update; the 32-bit point is the unquantized
    single-precision reference.
    Dashed lines show the corresponding 32-bit reference heat, and error
    bars are trajectory SEM within each screening condition.%
  }
  \label{fig:precision}
\end{figure*}

Figure~\ref{fig:precision}(a) shows the storage-precision
screen for the full-precision online checkpoint of seed 101.
With two-bit storage, very low heat is accompanied by a lower maximum
CNN softmax score, larger nearest-prototype distance, and strongly
reduced diversity.
Here low heat accompanies a narrower output distribution rather than
better generation. At 4 and 5 bits, the heat distributions
approach the 32-bit reference, and the 4-bit auxiliary quality summaries
are also close to their reference values.
We use the KS distance to describe differences in the heat
distributions and the auxiliary measures to assess quality and
diversity. The shared generation random numbers preclude using the
independent-sample KS $p$ value here.

The distinction between storage and training precision is pronounced
[Fig.~\ref{fig:precision}(b)].
With deterministic fixed-range rounding after every update, $J_{vh}$
remains identically zero at 4, 6, and 8 bits because typical updates are
below the grid spacing.  At 16 bits, the heat and auxiliary quality
summaries approach the 32-bit reference.  Stochastic rounding makes the
low-bit couplings nonzero and provides partial recovery, but the 8-bit
quality summaries still differ from the reference; the stochastic
16-bit result is again close to it.
Stochastic rounding partly recovers training at low precision. Even
with this recovery, 16 bits remain the lowest tested training precision
at which both heat and the auxiliary quality measures approach the
reference in this implementation.

\subsection{Learning rate and update timing}
\label{sec:numerical_controls}

\begin{figure*}[!t]
  \centering
  \includegraphics[width=\textwidth]{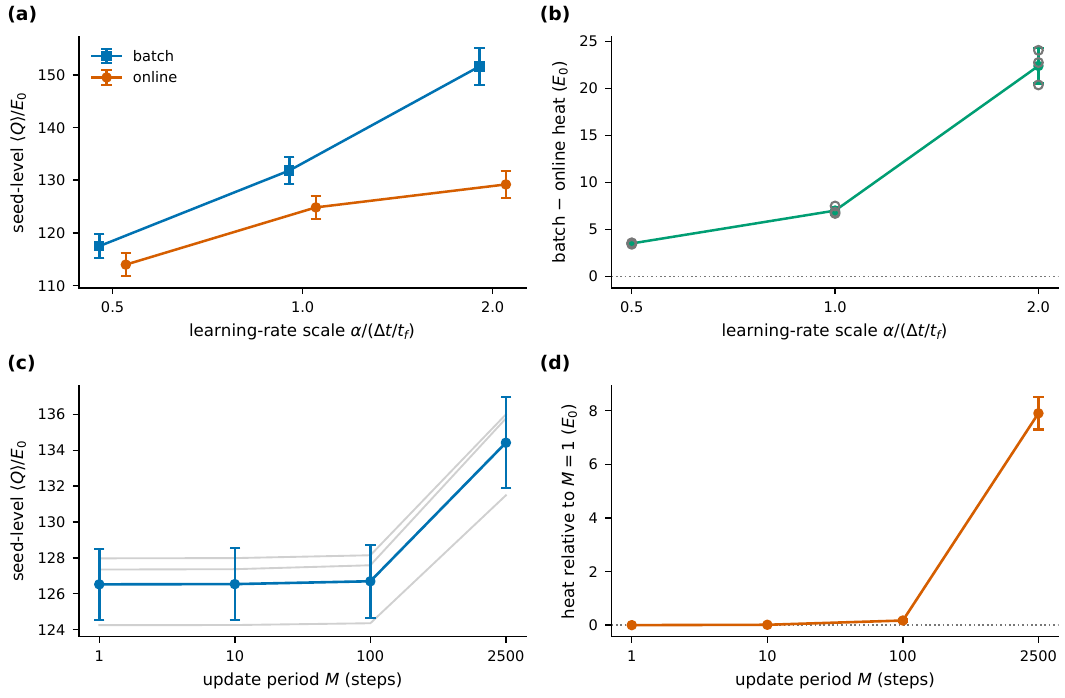}
  \caption{%
    \textbf{Optimization controls.}
    (a)~Seed-level heat for batch and online training across three
    learning-rate scales; error bars are 1 s.d. across seed means.
    (b)~Paired batch-minus-online heat; open circles show individual
    seed differences.
    (c)~Seed-level heat versus update period $M$; gray lines join matched
    seeds and error bars are 1 s.d. across seed means.
    (d)~Paired heat relative to $M=1$.
    The controls use three independently seeded pairs.%
  }
  \label{fig:optimization_controls}
\end{figure*}

Figure~\ref{fig:optimization_controls}(a,b) shows a three-seed
learning-rate control with $\alpha/(\Delta t/t_f)=0.5,1,$ and $2$,
$N_{\rm cycle}=100$, and $N_{\rm samp}=100$.
Online heat is lower in all three paired seeds at every scanned scale.
Both rules give their lowest measured heat at scale 0.5, where the
online models still release less heat. The difference therefore
persists across the tested settings. Because both minima lie at the
boundary of the scan, this control does not locate the optimal learning
rates. Comparing optima at matched generation quality would also
require quality measurements across the scan.

For the delayed-update control
[Fig.~\ref{fig:optimization_controls}(c,d)], local gradients are
accumulated for $M$ integration steps before being applied.
The fully online case is $M=1$, whereas $M=2500$ gives one accumulated
update per trajectory.
Across three seeds, $M=1$, 10, and 100 give similar heat, whereas the
fully accumulated case is consistently higher than $M=1$.
With $N_{\rm cycle}=80$ and $N_{\rm samp}=120$, accumulating gradients
over the full path produces models that release more heat during
generation than those trained with stepwise or moderately delayed
updates.

\subsection{Path likelihood and generation heat}
\label{sec:path_heat}

Why can models with similar training losses release different amounts
of heat? The loss measures reverse-path likelihood; the heat measures
energy released during generation. To relate them, consider a fixed
candidate parameter set $\parth$ and an observed noising path $\omega$.
Write $\mathbf{g}_{\theta,k}^{+}=\nabla\Vth(\xv_{k+1})$ and
$\mathbf{g}_{0,k}^{-}=\nabla V_0(\xv_k,t_k)$, where the uncoupled
reference potential $V_0$ includes the prescribed external forcing.
Expanding the Gaussian transition densities of
Eq.~\eqref{eq:loss_step} gives the exact discrete identity
\begin{align}
  \mathcal{R}_{\theta}
  &\equiv \ln\frac{P_0[\omega\mid\xv_0]}
                       {P_{\parth}[\tilde\omega\mid\xv_K]}
    \nonumber\\
  &= -\frac{\beta}{2}\sum_k\Delta\xv_k\cdot
       (\mathbf{g}_{\theta,k}^{+}+\mathbf{g}_{0,k}^{-})
    \nonumber\\
  &\quad+\frac{\beta\mu\Delta t}{4}\sum_k
       \left(\|\mathbf{g}_{\theta,k}^{+}\|^2
             -\|\mathbf{g}_{0,k}^{-}\|^2\right),
  \label{eq:path_action_ratio}
\end{align}
where $\beta=1/(\kbt)$ and the path densities are conditioned on their
respective starting states. The force-squared term is of order
$\Delta t$ per step and generally accumulates to a finite contribution
at fixed $t_f$. It must therefore be retained in the path-likelihood
ratio, which cannot be expressed in terms of heat alone for these two
models.

For generation under a frozen, time-independent potential, the released
heat instead follows directly from the energy change:
\[
  Q=-\int_0^{t_f}\nabla\Vth\circ d\xv
    =\Vth(\xv_0)-\Vth(\xv_{t_f}),
\]
where $\circ$ denotes the Stratonovich integral. We measure this heat
as the system relaxes in the trained potential. The measurement starts
after the parameters are set, excluding dissipation from writing the
parameters and the work needed to establish the potential.
The training objective is evaluated on externally generated noising
paths, whereas $\avgQ$ averages over paths generated by the frozen
trained model. The averages are over different ensembles, so minimizing
the training loss does not by itself minimize generation heat.
Online and batch training evaluate the same fixed-parameter objective
along different finite-step optimization trajectories
[Eq.~\eqref{eq:online_batch_expansion}], and can select solutions with
similar path loss but different generation heat.

Evaluating the saved models with fixed parameters makes this
distinction concrete.
On the 30 additional noising paths, the normalized losses remain
numerically close after subtracting the common squared-increment term,
with a slightly lower loss for online training. The similar losses
are therefore not simply a consequence of the large common offset.

The energy decomposition locates the heat difference at the generated
endpoints. The two models start from nearly equal mean energies.
Online-generated states have less negative coupling energy and smaller
positive onsite energy than batch-generated states, evaluated in their
respective learned potentials. The coupling-energy change dominates,
leaving a higher final total energy and hence a smaller energy release.
The online checkpoints also have smaller coupling Frobenius norms in
both coupling blocks in all five pairs.
Evaluating both potentials on both endpoint ensembles distinguishes
changes in the potential from changes in the sampled states. On
either ensemble, the online potential gives a smaller mean energy
drop. In either potential, the online endpoint ensemble also gives a
smaller drop. These comparisons connect the lower heat to changes in
both the learned potential and the states reached during finite-time
generation, while the path losses remain close.

\subsection{Noise structure and physical implementation}

The structure of the errors helps explain the different noise responses.
Independent zero-mean errors can partly cancel over many steps,
consistent with the weak heat response at low formed-update noise.
This averaging also occurs when stepwise estimates are accumulated
before updating; it is not specific to the online rule.
Offsets and slowly correlated errors persist across steps, limiting
what this averaging can remove and making calibration more important.

The response curves in Secs.~\ref{sec:noise} and~\ref{sec:quant}
show how each perturbation affects learning and generation. They also
separate two precision requirements: retaining the behavior of a trained
model when its couplings are stored, and resolving small increments
when those couplings are updated. The thresholds depend on the task
and protocol. We assess storage through heat and auxiliary quality
measures with parameters fixed, and training through fixed-range
quantization after every update. Relating the dimensionless error
amplitudes to device parameters requires a model of the chosen platform.

The locality of Eq.~\eqref{eq:online_J} suggests a route to physical
implementation.
Each coupling gradient is a symmetric sum of residual--state
correlations measured at one timestep; computing it requires no
backpropagation through the trajectory.
This distinguishes the present setting from analog in-memory neural
network training methods designed to handle asymmetric device noise in
crossbar arrays~\cite{Gokmen2016,Nandakumar2020,Gokmen2020}.
Possible substrates include coupled electrical, mechanical, or
superconducting stochastic systems~\cite{Melanson2025,Dago2021}.
For each platform, the relevant questions are how to measure the local
residuals and apply coupling increments with controlled noise, drift,
and precision.

\FloatBarrier

\section{Conclusions}
\label{sec:conclusions}

We train a generative Langevin computer one integration step at a time,
updating its couplings with a symmetric sum of local residual--state
correlations. Across five paired digital simulations, online and
trajectory-batch training reach similar validation losses on fixed
paths, while the online models release less heat on average during generation
in every pair. Endpoint diagnostics connect this difference to the
learned potentials and the states they generate. Lower heat accompanies
reduced sample diversity. Independent zero-mean errors in the formed
updates and coherent errors in the residuals have markedly different
effects on heat, and storing trained couplings requires less precision
than resolving deterministic training updates. For the tasks and
protocols studied here, these simulations show how update timing, noise
structure, and precision shape local learning in generative
thermodynamic computers.

\begin{acknowledgments}
This work was supported by the National Key Research and Development
Program of China under Grant No.\ 2024YFA1611003, the Fundamental
Research Funds for the Central Universities (XJ2026002701), the Natural
Science Foundation of Fujian Province (Grant No.\ 2026J0011623), and
the 111 Project 2.0 (Grant No.\ BP0820038).
\end{acknowledgments}

\section*{Data Availability}

There are no publicly available research data or software supporting this
manuscript. Requests for further information or data should be sent to the
authors.

\clearpage
\bibliography{references}

\end{document}